\documentclass[aps,showpacs,eps,preprint]{revtex4}
\usepackage{graphicx}
\usepackage{colordvi}
\usepackage{amssymb}
\usepackage{amsmath}
\usepackage{bbold}
\usepackage{epsf}
\usepackage{mathrsfs} 
\usepackage{float}
\usepackage{subfigure}
\begin{document}
\def \beq{\begin{equation}}
\def \eeq{\end{equation}}
\def \bea{\begin{eqnarray}}
\def \eea{\end{eqnarray}}
\def \bem{\begin{displaymath}}
\def \eem{\end{displaymath}}
\def \P{\Psi}
\def \Pd{|\Psi(\boldsymbol{r})|}
\def \Pds{|\Psi^{\ast}(\boldsymbol{r})|}
\def \Po{\overline{\Psi}}
\def \bs{\boldsymbol}
\def \bl{\bar{\boldsymbol{l}}}

\title{A Green's function approach to  transmission of massless Dirac fermions in graphene through an array of random scatterers} 
\author{Neetu Agrawal (Garg)$^1$, Sankalpa  Ghosh$^2$ and  Manish Sharma$^1$}
\affiliation{$^1$Centre for Applied Research in Electronics, Indian Institute of Technology Delhi, New Delhi-110016, India}
\affiliation{$^2$Department of Physics, Indian Institute of Technology Delhi, New Delhi-110016, India}

\begin{abstract}

We consider the transmission of massless Dirac fermions through an array of short range scatterers which are modelled as randomly positioned $\delta$- function like potentials along the $x$-axis. We particularly discuss the interplay between disorder-induced  localization that is the hallmark of a non-relativistic system and two important properties of such massless Dirac fermions, namely, complete transmission at normal incidence and periodic dependence of transmission coefficient on the strength of the barrier that leads to a periodic resonant transmission.  This leads to two different types of conductance behavior
as a function of the system size at the resonant and the off-resonance strengths of the delta function potential. 
We explain this behavior of the conductance in terms of the transmission through a pair of such 
barriers using a Green's function based approach. The method helps to understand such disordered transport in terms of well known optical phenomena such as Fabry Perot resonances. 

\end{abstract}
\date{\today}
\maketitle

\section{Introduction}
The remarkable transport properties of graphene \cite{Geim, review} are primarily due to the fact that 
under ambient conditions the charge carriers are massless Dirac fermions with definite chirality. Such electrons get 
differently scattered by a potential barrier \cite{KNG06} as compared to non-relativistic electrons  in other conventional semiconductor or metal . 
This led to a number of transport anomaly in the graphene. The expression of  conductivity of ballistic graphene is  remarkably different \cite{kat, two}  from that of a non-relativistic electron  and the  minimum conductivity is given $\frac{4e^2}{h}$ for undoped graphene. The prefactor $4$ is due to contribution from two sublattice degrees of freedom and two 
valley degrees of freedom. 

This prediction was verified experimentally \cite{Heersche, miao}.
More remarkably, addition of impurities to  pristine monolayer graphene leads to the conductivity enhancement at the Dirac point, where as 
addition of such impurities away from the Dirac point leads to a supression of conductance \cite{titoveuro,  Xiong, titovprl}. 
A general theory to understand the transport of Dirac fermions in presence of such isolated impurities was also developed \cite{Mirlin}
 A more complete theory that takes into account the effect of disorder as well as interaction effect was also constructed. 
\cite{Rossi} that gives better agreement with the transport measurement.  
Such impurity scatterers could be vaccancies, adsorbed atoms, molecules, or impurity clusters, \cite{wehling}\cite{schedin}, or hydrogen atoms controllably added to the surface \cite{elias} or metallic islands deposited on graphene surface \cite{kessler}.The difference between the influence of short range scatterers and long range Coulomb impurities on the transport of massless Dirac fermions 
was also studied \cite{Nomura}. 
Theoretical progress was also made to understand the nature of transport in presence of correlated disorder \cite{Rossi2}.

For a better theoretical understanding of these transport properties of massless Dirac fermions in graphene, it 
must be compared with the conventional charge transport in disordered condensed matter system  which 
is principally understood in the frame work of the 
 pioneering work of Anderson\cite{anderson} and subsequent developements \cite{Ramakrishnan} in this direction.
Particularly important in the context of graphene which is a two-dimensional atomic crystal the prediction of
 scaling theory of localization \cite{scaling}.  The scaling theory predicts that below two dimension any amount of disorder can localize all the states.
One of the implications of this scaling theory is that  conductace G approaches zero as the sample size L goes to infnity for a disordered one dimensional system and  such decay is exponential in nature \cite{oned}. These predictions are based on the 
fact that charge carriers are non-relativistic in nature and obeys Schr\"odinger equation. Thus it is extremely important to study the revision of above well established properties of non relativistic electrons for the case of massless chiral  dirac fermions that dominates the transport properties of graphene. 

The transmission of such massless dirac fermion  through one dimensional potential barriers 
demonstrates two fundamentally different behavior as compared to the similar transmission of 
non relativistic electrons \cite{KNG06}. First of all, they Klein tunnel through such barrier which implies full transmission at normal incidence. Secondly the transmission prbability periodically oscillates with the varying strength 
of the barrier.  This particular property which was already implicit in the transmission expression given in \cite{KNG06} was 
more clearly demonstrated in the ref. \cite{titoveuro} by considering transmission through short range scatterers 
approximated as delta function potential. 

The above mentioned exotic properties of transport electrons  led to a large body of work devoted to the study of ballistic progation of two dimensional 
massless Dirac fermions in graphene through various one dimensional superlattices made out of scalar and vector potentials \cite{Chpark1, Chpark2, Masir, peeter, Arovas, magbarrier}.
Experimentally also certain superlattice structures were imposed on monolayer graphene \cite{Expt}. 
Relatively much less attention has been paid \cite{Onedirac, muller} to similar studies through one dimensional array of disordered or impurity potential. In this paper we study the transmission of such massless dirac fermions  through a one dimensional arrangements of short range impurities modelled as delta function potetials on random locations using a Green's function based technique. We obtain an analytical  expression for the transmission and zero temperature conductance 
through such one dimensional arrangement of delta function scatterers. 

Using this theoretical framework we show 
that the transmission and conductance in the presence of array of disorder potentials can be understood in terms 
of the transmission through a double barrier structure consisting of such delta function like potentials. 
One of our main results that the conductance properties through such barriers can have two type of behaviors
as a function of the system size. If the strength of the each delta function barrier is close to its resonant values \cite{titoveuro} then there is perfect transmission through each barrier and the conductance  does not change 
significantly as the system size increases. On the otherhand if the strength of such delta function scatterers is well off 
from the resonant value, the conductance shows a algebraic decay as a function of the size of the system with an exponent which is evaluated to be close to $0.5$. As our results shows this happens over a wide range of the strength as well as the 
mean separation between two such successive scatterers. The Green's function method that we use is quite general and can also be used to reporduce well known results \cite{peeter, Masir} for ordered superlattice  such as Kronig Penny systems and their different variants \cite{Vrinda}.

The paper has been organized in the following manner. In various subsections of section \ref{theory} we develope the theoretical framework of the Green's function method and derived the expression of transmittance and conductance. In Section \ref{results} we first 
explain the single and two barrier transmission and explain how the $N$ barrier transmission can be understood in terms 
of successive two barrier transmission. We then present the results for transmission through $N$ such barriers, first when 
the positions of such barrier is randomly located, but all having equal strength, and then with both position as well as strength randomized. We conclude by pointing out a possible generalization of our technique and implications of our result in Section \ref{concl}.

\section{Green's function and the Transmission}\label{theory}

In this section we first obtain the free particle Green's function for massles Dirac fermions and determine the solution 
in presence of short range scatterers modelled as an array of the delta function potentials. We then analyse how these solutions can be understood 
in terms of multiple scattering processes takes place between two such barriers. We shall finally obtain the expression for the transmittance through $N$-such 
barriers with arbitrary position and strength.  

\subsection{ Wave functions of massless dirac fermions in terms of free particle Green's function}
The charge carrier in Graphene under ambient condition behave like two dimensional masssless Dirac fermions 
 \cite{{KNG06}}. For such charge carriers with energy $E$, the stationary solutions are obtained from the following Dirac-Weyl equations
\beq  -i\hbar v_F\vec{\sigma}\cdot\vec{\nabla}
\psi(x,y) + V\mathbb{1}\psi(x,y) = E\psi(x,y) \label{eq1} \eeq 
Here $ v_F \approx 10^6ms^{-1}$ is the Fermi velocity  and $\vec{\sigma}=\sigma_{x} \hat{x} + \sigma_{y} \hat{y}$.  
Here $\psi(x,y)$ is a two component pseudo-spinor  where the pseudospin refers to the sublattice degrees of freedom. We take $V(x,y)=V(x)$ such that translational invariance along $y$-directions leads to $\psi(x,y) = \left(\begin{matrix} \psi_1(x) \\ \psi_2(x) \end{matrix}\right)e^{ik_y y}$.  Rescaling lengths and energy like quantities by 
Fermi wavelength $ \lambda_F=\frac{2 \pi}{k_{F}}$ and $\hbar v_F/l_s$, we introduce dimensionless variables  $\bar{x}=x/ \lambda_F$, $\bar{k}_{x,y} = k_{x,y} \lambda_{F}$, $\bar{\epsilon}=E \lambda_F/\hbar v_F$, and $\bar{v}=V\lambda_{F}/\hbar v_F $. 
Substituting the $y$-translationally invariant solutions in   Eq. (\ref{eq1}) and multiplying by $\sigma_{x}$, in terms of the dimensionless variables mentioned, we get the effective one dimensional equation 
\beq i\frac{d}{d\bar{x}}\psi(\bar{x})+\left(\begin{matrix} -i\bar{k_y} & \bar{\epsilon} \\ \bar{\epsilon} & i\bar{k_y} \end{matrix}\right)\psi(\bar{x}) = \bar{v}(x)\sigma_x\psi(\bar{x}) \label{eff1D} \eeq
We are interested in seeking the nature of transmission of such massless Dirac fermions through a potential of the form 
\beq  \bar{v}(\bar{x})=\sum^{N}_{l=1}\lambda_{l}\delta(\bar{x}-{\bar{x}}_{l}) \label{potn} \eeq
The above potential is a series of delta functions that are randomly positioned within a length (say) L, such that the end points are held fixed at $ x_1 = 0$ and $x_N = L$. A delta function is placed at each of the edges $ x_1$ and $x_N$ while the number in between the edges can be varied. Additionally, the strength $\lambda_l$ of each of the delta function can also be varied. Thus we are studying the transmission of two dimensional Dirac fermions through an one dimensional potential that mimics the effect of a set of random short range scatterers on a line. 
If the range of the potential of isolated impurities is much smaller than the fermi wavelength of electrons ( which is theoretically infinity at the Dirac point), but larger than the carbon-carbon bond 
length in monolayer Graphene, the scattering potential of an isolated impurity can be modelled as a delta function. 

Here we describe the main steps of the Green's function based method that we adopted to calculate such transmission. The method particularly takes care of the massless ultra relativistic nature of such fermions. 


The eq.(\ref{eff1D}) is an inhomogenous differential equation whose full solutions can be written as a  sum of the complementary fnction and particular integral of which the later can be expressed in terms of the Green's function.
The resulting solution can be written as 
\beq \psi(\bar{x}) = Ae^{i\bar{k}_x\bar{x}} +  Be^{-i\bar{k}_x\bar{x}} + \int^{\infty}_{-\infty}d\bar{x}^{'}G(\bar{x},\bar{x}^{'})^{\sigma}\sigma_x\bar{v}(\bar{x}^{'})\psi(\bar{x}^{'}), \label{totsoln} \eeq 
where the Green's function $G(\bar{x},\bar{x}^{'})^{\sigma}$ is a $2\times 2$ matrix given as 


\beq G(\bar{x},\bar{x'})_{ \bar{x} > \bar{x'}}^{\sigma} = e^{i\bar{k}_{x}(\bar{x}-\bar{x'})} G^{\sigma} \label{Greenf} \eeq 
The pseudopsin dependent part of the Green's function is given by 
 \beq G^{\sigma} = \frac{-i}{2\cos\phi}\left(\begin{matrix}e^{-i\phi} & 1 \\ 1 &e^{i\phi} \end{matrix}\right) \label{spingreen} \eeq 

Similarly \beq G(\bar{x}, \bar{x'})_{ \bar{x} < \bar{x'}}^{\sigma} = \frac{-i}{2\cos\phi}  e^{-i\bar{k}_{x}(\bar{x}-\bar{x'})} \left(\begin{matrix}-e^{i\phi} & 1 \\ 1 & -e^{-i\phi} \end{matrix}\right) \eeq

In the above solutions $A = A_1\left(\begin{matrix} 1 \\ e^{i\phi} \end{matrix}\right)$, $B = B_1\left(\begin{matrix} 1 \\ -e^{-i\phi} \end{matrix}\right)$, $A_1$ and $B_1$ are constants. 

 $\phi = tan^{-1} \bar{k}_y/\bar{k}_x$ with $\bar{k}_F^2 = \bar{k}_{x}^2 + \bar{k}_{y}^2$

Substituting the form of potential given in (\ref{potn}) in  the solution (\ref{totsoln}) we get the stationary solution at any spatial point $x$ as 
\beq \psi(\bar{x}) = Ae^{i\bar{k}_x\bar{x}} +  Be^{-i\bar{k}_x\bar{x}} + \sum^{N}_{l=1}\lambda_{l}G(\bar{x},\bar{x}_l)^{\sigma}\sigma_x\psi(\bar{x}_l) \label{redsoln} \eeq

\subsection{Multiple scattering processing and the solution at N-th barrier} 
The above solution (\ref{redsoln}) implies that  we need to determine $\psi(\bar{x}_l)$ at the locations for each delta function, namely at $l = 1$ to N to get the solutions at any spatial point $x$. This is due to the fact that  
they are the only scattering centers that can change 
the incident amplitude. It may be pointed out that above method of calculating the amplitude through the solution (\ref{totsoln}) 
 can be implemented for any other form of potential in one and higher dimension as well. However in the presence of delta function like scattering centers the integral collapses into analytically tractable series of  the series  given in  (\ref{redsoln}).
Finally after having the solution at each point of space we can calculate the ratio of transmitted current density and 
incident current density to yield  the transmission coefficient through such series of scatterers.  
However before proceeding to determine the expression for the transmission coefficient using this method we shall discuss briefly some issues that are relevant to the determination of 
the value of $\psi( \bar{x}_{l})$  at the location of the dirac delta function.

\begin{figure}[t]
\begin{center}
\centerline{\epsfxsize 5cm \epsffile{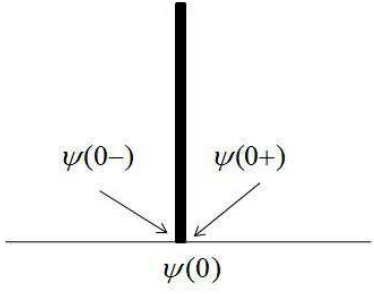}}
\end{center}
\caption{Schematic display of the matching condition in presence of delta function barrier.}
\label{schema}
\end{figure}

For a non-relativistic spinless particle obeying one dimensional Schr\"odinger like equation, the solution at the location at the 
delta function is obtained by taking the average of the left and right hand limit, namely 
\beq \psi(0) = \frac{1}{2} [ \psi( 0 +) + \psi(0-)] \nonumber \eeq 
This situation is schematically pointed out in Fig. \ref{schema}. 
However as pointed out in a number of works \cite{ MS, Roy, mckellar, peeter} this procedure cannot be used in determining the solution at the location of the delta function for the corresponding relativistic problem obeying the one dimensional 
dirac equation which is a first order differential equation. The issue is resolved by matching the two component spinorial 
wavefunction on the right and left side of a finite width barrier and then finding out of the form of the transfer matrix 
by allowing such fnite width barrier to approach delta function limit.  

Such matching condition  gives is 
\beq \left(\begin{matrix}\psi_1(0+)\\ \psi_2(0+)\end{matrix}\right) = S \left(\begin{matrix}\psi_1(0-)\\ \psi_2(0-)\end{matrix}\right) \label{smat}\eeq

\begin{center}
\framebox[1.5\width]{$S = \left(\begin{matrix} \cos\lambda &- i\sin\lambda \\- i\sin \lambda & \cos \lambda\end{matrix}\right)$}
\end{center}

We note that S is a periodic function of $\lambda$ and that S = $\pm I $ for $\lambda = n\pi$. This the special situation for which the barriers behave as if they are transparent to the incident charge carriers.

Using the above form of the matching matrix, the wavefunction at the position of the delta function is expressed as the following linear combination:
\beq   \psi(\bar{x}_l+)+\psi(\bar{x}_l-) = \lambda \cot\frac{\lambda}{2}\psi(\bar{x}_l)  \label{matchcondn} \eeq 
$\lambda \neq 2n\pi, n= 0,1,2,3 \cdots $
\cite{Roy}. The corresponding method  the case of $\lambda = n \pi$ where the matching matrix is a unity matrix and leads to full transmission 
through the delta function scatterers via the matching condition (\ref{smat}) and will be described later. 
The above method will now be implemented first to determine to evaluate $\psi(\bar{x}_l)$ for all l values, namely for  
 $l = 1$ to$N$.

To calculate the transmission through such delta function barriers we need to evaluate the wavefunction at the last such barrier which is the $N$-th one, namely $\psi(x_{N})$ .  According to eq. \ref{redsoln} this can be written in terms of   
$\psi(\bar{x}_{N} \pm)$.  Now, 
\bea \psi(\bar{x}_N+)  & = &  Ae^{i\bar{k}_x\bar{x}_N} +  Be^{-i\bar{k}_x\bar{x}_N} + \sum^{N}_{l=1}\lambda_{l}G(\bar{x}_N+,\bar{x}_l)^{\sigma}\sigma_x\psi(\bar{x}_l)  \nonumber \\
& = & (A + S_{-}(N)) e^{i\bar{k}_x\bar{x}_N} +  Be^{-i\bar{k}_x\bar{x}_N}  \label{psiplus}\eea
Here we  used the scattering matrix $S_{-}(N)$ defined as 
\beq S_{-}(N) = \exp(-i\bar{k}_{x} \bar{x}_{N})
\sum_{l=1}^{N} \lambda_{l} G(x_{N}+, x_{l})^{\sigma}\sigma_{x} \psi ( \bar{x}_{l}) \label{Sn} \eeq 
which gives us the modification of the amplitude in the forward propagating wave $A$ due to scattering by all the 
$N$ barriers.

Similarly the amplitude at the immediate left of the $N$-th barrier 
\beq \psi(\bar{x}_N-)   =   Ae^{i\bar{k}_x\bar{x}_N} +  Be^{-i\bar{k}_x\bar{x}_N} + \sum^{N}_{l=1}\lambda_{l}G(\bar{x}_N-,\bar{x}_l)^{\sigma}\sigma_x\psi(\bar{x}_l) 
\label{psiminus}\eeq

Using the above expressions and the relation (\ref{matchcondn}) a straightforward calculation 
can determine 

\beq \psi( \bar{x}_{N}) =\frac{ \sin \frac{\lambda_{N}}{2}}{\frac{\lambda_{N}}{2}} \exp( i \bs{\sigma} \cdot \hat{x} \frac{\lambda_N}{2})  \left[\left(A+S_{-}(N)\right)e^{i\bar{k}_x\bar{x}_N} +  Be^{-i\bar{k}_x\bar{x}_N}\right] \label{psiN} \eeq 

The above expression can be understood as follows. The first term $\frac{ \sin \frac{\lambda_{N}}{2}}{\frac{\lambda_{N}}{2}}$   which provides the envelope of the propagating wave through the delta function 
barrier is due to the Fraunhofer diffraction  of the wave function by the delta function barrier. The second term
is due to that because of the linear dispersion of the massless Dirac fermions, the delta function potential barrier 
directly changes the phase of the pseudospinor by changing the $x$ component of the wave vector. This fact is 
expressed  by the explicit presence of the pseudopsin rotation operator, where the angle through which the pseudospin rotation takes place is directly proportional to the strength of the delta function potential. The third term is the modification 
of the free particle wave function due to the multiple scattering from the $N$ delta function barriers. It may be noted 
because of the presence of $S_{-}(N)$, the scattering matrix the determination of  $\psi(\bar{x}_{N})$ requires the determination of $\psi ( \bar{x}_{l})$  for $l=1$ to $N$. The expression for $\psi(\bar{x}_{l})$ can again be obtained 
in the same way as in the case of $\psi(\bar{x}_{N})$, yielding

\beq \psi(\bar{x}_l)= \frac{\sin \frac{\lambda_{l}}{2}}{\frac{\lambda_l}{2}}\exp( i \bs{\sigma} \cdot \hat{x} \frac{\lambda_N}{2})  
 \left\{\rho_l[A-S_{-}(N)]e^{i\bar{k}_x\bar{x}_l} +\bar{\rho}_lBe^{-i\bar{k}_x\bar{x}_l}\right\} \label{psixl} \eeq
Since the right hand side of the above equation also contains $\psi(\bar{x}_{l})$ to determine the the scattering matrix 
$S_{-}(N)$, it has to be solved self consistently to get the solution $\psi(\bar{x}_{l})$

The term $\rho$ which is given as
\beq  \rho_l =  I + \sum_{j=l+1}^{N}\eta_{j_1l} + \sum_{j_2 = l+2}^{N}  \sum_{j_1 = l+1}^{j_2-1}\eta_{j_1l}\eta_{j_2j_1} +  \sum_{j_3 = l+3}^{N} \sum_{j_2 = l+2}^{j_3-1}  \sum_{j_1 = l+1}^{j_2-1}\eta_{j_1l}\eta_{j_3j_2}\eta_{j_2j_1} +  \cdots \label{rhoform}  \eeq

Where
\beq \eta_{jl} = \lambda_{j} \frac{\sin \frac{\lambda_{j}}{2}}{\frac{\lambda_j}{2}} \exp( i \bar{k}_{x} ( \bar{x}_{j} - \bar{x}_{l})) 
[G(\bar{x}_{j}, \bar{x}_{l})^{\sigma}_{\bar{x}_l<\bar{x}_j} - G(\bar{x}_{l}, \bar{x}_{j})^{\sigma}_{\bar{x}_j>\bar{x}_l}] \sigma_{x}\exp( i \bs{\sigma} \cdot \hat{x} \frac{\lambda_{j}}{2}) \label{etamat}
\eeq
and $\bar{\eta}_{jl}$ is just the complex conjugate of the $\eta_{jl}$. 

Thus $\eta_{jl}$ transfer the amplitude from $j$ the delta function to the $l$ delta function due to the scattering at the $j$-th delta function. 
The series $\rho_{l}$ indicates how the amplitude is transferred to the $l$-th barrier from all the barriers that succeeds it as one goes from left to write 
through the multiple scattering processes of various order. 

The first term is an identity matrix and indicate the zeroeth order process, namely when 
$l=N$, the final barrier as was the case for the expression given in (\ref{psiN}). The  second term refers through a single scattering process by which the 
amplitude is transferred from all the barriers right to the $l$-th barrier to the $l$-th barrier through a single scattering. The third term refers to the processes 
where such amplitude transfer takes place through two successive scatterings and hence a second order process. 

The scattering matrix $S_{-}(N)$ can be now be self consistently determined  by substituting eq. \ref{psixl} into eq. \ref{Sn} which can be written in terms 
of pre determined quantities such that 
\beq S_{-}(N)=\left[I+\sum^{N}_{l=1}\chi_l\rho_l\right]^{-1} \sum^{N}_{l=1}\chi_l\left(\rho_lA +\bar{\rho}_lBe^{-2i\bar{k}_x\bar{x}_l}\right)  \label{S} \eeq
where 

\beq \chi_{l}= \lambda_{l} \frac{\sin \frac{\lambda_{l}}{2}}{\frac{\lambda_l}{2}} G^{s}\sigma_{x}\exp( i \bs{\sigma} \cdot \hat{x} \frac{\lambda_{l}}{2}) \eeq 
 where $G^{s}$ is the spin compoent  for  the Green's function . 


\subsection{Calculation of Transmission coefficient}
In our present theoretical framework we are assuming the massless dirac fermion to incident from the left side of the first barrier and after $N$-the barrier it will again freely propagate 
towards the right. The structure of $S_{-}(N)$ and $\rho_{l}$ implies this assumption. 
To calculate the transmission coefficient in the presence of (say) N number of barriers we need to evaluate the wavefunction for $\bar{x}<\bar{x}_1$ and for  $\bar{x}>\bar{x}_N$. For   $\bar{x}>\bar{x}_N$, all the scattering centres which lie behind $\bar{x}_N$ contribute. Also since there is no wave coming from the right side 
 we set $B=0$. Hence

\beq \begin{matrix} \psi(\bar{x}) = [A+S_{-}N]e^{i\bar{k}_x\bar{x}}   &   &  &  \bar{x}>\bar{x}_N  \end{matrix} \nonumber   \eeq 

The transmittance $T$ can be obtained as the ratio of the transmitted probability density and the 
incident probability density is obtained as:
\beq T=\left|t\right|^2  =  [A+S_{-}N]^{+}[A+S_{-}N]/A^{+}A \nonumber \eeq

Setting the condition $B=0$ in  the expression in the expression (\ref{S}) of the scattering matrix $S_{-}(N)$ 
\beq A+S_{-}(N)=\frac{1}{det\left[I+\sum^{N}_{l=1}\chi_l\rho_l\right]}cof\left[I+\sum^{N}_{l=1}\chi_l\rho_l\right]A \nonumber \eeq 
Using the relation \cite{reading}
\beq  cofactor\left[I+\sum^{N}_{l=1}\chi_l\rho_l\right]A = A \nonumber \eeq
we obtain 
\beq T= \left|\frac{1}{det\left[I+\sum^{N}_{l=1}\chi_l\rho_l\right]}\right|^2  \label{transformula}  \eeq
It may be mentioned for $\lambda = n \pi$, $T=1$ since the matching matrix is an identity matrix. 
The above expression provides us for a given angle of incidence $\phi$, 
the transmittance through a randomly positioned $N$ delta function like barriers
whose strength $\lambda_{l}$ can also vary from point to point. The expression for the two terminal  conductance 
can now be obtained by suitably integrating the above expression for all possible angle of incidenc 
using the expression $G = G_0\int_{0}^{\pi/2}T(\lambda,\phi)\cos\phi d\phi$. Here, 
$G_0 = 4E_FL_ye^2/(v_Fh^2)$ and $L_y$ is the width of the system.

For the current problem the expression for two terminal conductance in the presence of $N$ random 
delta function like impurities on a line
will be 
\beq G = G_0\int_{0}^{\pi/2}\left|\frac{1}{det\left[I+\sum^{N}_{l=1}\chi_l\rho_l\right]}\right|^2\cos\phi  d\phi  \label{condexp} \eeq
It maye be mentioned that the same problem can also be solved with the transfer matrix method. But the interpretation of the finally transferred amplitude
in terms of a multiple scattering process at various orders comes out explitly in the Green's function framework. This provides a way to understand scattering through $N$ random barriers 
with the help of scattering through two barriers. 
A connection between the Green's function method 
and the transfer matrix method for non-relativistic electrons is provided in ref. \cite{reading}. The method is also similar in spirit to the invariant embedding 
approach used in radiative transfer processes \cite{radiative}. 

In the subsequent section we use this expression to evaluate the transmittance and conductance under various conditions 
and analyze these results.

\section{ Results and Discussions}\label{results}
In this section we shall shall use the expression (\ref{transformula})  and (\ref{condexp}) to calculate the transmission 
and  resulting conductances through such various combinations of the delta function like barriers. We shall first analyze the 
case of the transmission through a single delta function like barrier and two delta function like barrier to understand some 
peculiarity of one dimensional transmission of two-dimensional massless Dirac fermions in Graphene through such 
single and double barrier structure as compared to a corresponding non relativistic problem.  We explain how these 
peculiarities are expected to change the localization properties of such two dimesional massless Dirac fermions. Subsequently we shall present our result for transmittance and conductance through a random array of such delta 
function barriers over a wide range of strength as well as delsity of such barriers and discuss the nature of localization 
of such massless Dirac fermions in presence of short range scatterers. 

\subsection{Single delta function barrier}
The transmittance of two dimensional massless Dirac fermion of energy $E$   \cite{KNG06} through
a one dimensional potential barrier of height $V_{0}$ and width $D$ 
such that  $\left|V_0\right|>>E$ is given by \cite{KNG06} the well known expression

\beq T = \frac{\cos^2\phi}{1-\cos^2(q_xD)\sin^2\phi} \label{Tformula} \eeq
where $ q_x = \sqrt{\frac{V_0^2}{\hbar^2v_F^2}-k_y^2}$ and $k_y = \frac{E\sin\phi}{\hbar v_F}$
The above expression has two important difference with the corresponding non-relativistic problem. 
At normal incidence for $\phi=0$, $T=1$ due to Klein tunneling. Also everytime the condition $q_{x}D = n \pi$ is 
satisfied $T=1$, showing resonant transmission with $\pi$ periodicity. 

The delta function limit of such square barrier is taken by allowing $ V_{0} \rightarrow \infty$ and $D \rightarrow 0$ , such 
that $ \frac{V_{0}D}{\hbar v_{F}} \rightarrow \lambda$ and the resulting transmission can be obatined as 

\beq  T = \frac{1}{1+\sin^2\lambda \tan^2\phi} \nonumber \eeq
The above expression can be directly obtained from the expression (\ref{transformula}) after setting 
in the equation \ref{potn}
\beq  \bar{v}(\bar{x})=\lambda \delta(\bar{x}) \eeq

A comparison with the corresponding results for a non-relativistic problem shows that in the case of the later the transmittance decays with the increasing strength of the delta function barrier. 

\subsection{Two delta function barriers}
Next we study the case of transmission of such massless dirac fermions through two such delta function barriers which 
are separated by a distance $\bar{d}$ ( in dimensionless form). The scattering potential takes the form 
\beq v(\bar{x})= \lambda_{1}\delta(\bar{x} +\frac{\bar{d}}{2})  + \lambda_{2} \delta(\bar{x} - \frac{\bar{d}}{2})
\nonumber \eeq 
Such a potential structure forms a Fabry perrot cavity for the massless Dirac fermions and the transmission through such 
cavity shows Fabry Perrot resonances \cite{Fabry}. 
According to (\ref{transformula}), for $\lambda_{1}=\lambda_{2}=\lambda$, the transmittance through such double barrier structure is given by  
 \beq T = [1+tan^2\phi(\cos \bar{k}_x\bar{d}\sin2\lambda-2\sin\bar{k}_x\bar{d}\sin^2\lambda/\cos\phi)^2]^{-1} \nonumber \eeq
The above expression clearly demonstrates that perfect transmission takes place respectively when $\phi=0$, $\lambda = n \pi$ and $\cos \phi=\tan \bar{k}_{x} \bar{d} \tan\lambda $. The first two conditions respectively correspond
to the Klein tunneling through such barrier and resonant transmission as the strength of barrier is varied. Both features are  single 
barrier transmission feature and reoccurs when two such barriers are placed side by side. 
The third condition refers to the Fabry Perrot resonance condition is a purely double barrier feature and occurs due to  multiple reflections from the two-barrier structure. 

In section \ref{theory}  we developed the expression for transmission through 
$N$ such delta function barriers, by writing the scattering matrix $S_{-}(N)$. This in turn  is determined by the series expression of 
the $\rho_{l}$ matrix given in (\ref{rhoform}) whose various terms depicting the different order multiple scattering processes.
Each such term  can be written as a product of $\eta$ matrices given by the expression (\ref{etamat}) which depicts the transfer of amplitude from one barrier to 
the other. Thus the features through  N delta functions  barriers, that have the strength, but located 
randomly on a line can be understood through in terms of multiple two barrier transmissions.  This is one of the main results of the current work. 

The above statement can be well explained using theory of Fabry-Perrot resonances of massless Dirac fermions \cite{Fabry}
through such double barrier structure.  If the strength of the delta function barriers are equal and their respective transmission and reflection amplitude is given by $t$ and $r$ 
the multiple scattering processes that involves any two such barriers is a linear superposition of terms like 
$tt, ttr^2\exp(i \zeta), \cdots, tt\exp(i2(n-1)\zeta)$. 
In all these terms $t$ is independent of the separation between two such barriers 
where as the phase change $\zeta$ that is acquired at each scattering between two barriers is dependent on the separation between 
such barriers. Since for massless Dirac fermions $t^{2}( \lambda)  = t^{2}( \lambda + n \pi)$, 
 such a linear superposition is also going to be $\pi$-periodic in the strength 
of the barrier $\lambda$. The same argument can be extended to the multiple scattering process that involves transfer of amplitude from 
$j$-th barrier to the $i$-th barrier through any number of intermediate barriers. Such a transmission can be wriiten in terms of the product 
of the $\eta$ matrices defined in (\ref{etamat}) each of which is going to be periodic in terms of the strength of the barrier. However now these barriers being positioned randomly on a line, the phase change induced by each of these pair of $\delta$ function barriers is random. Thus the total transmission 
through such barriers will be given by  a sum of terms whose amplitude is periodic in the strength of the potential $\lambda$, but phase is random.
The resulting transmission will therefore show periodicity as a function of the barrier strength $\lambda$, unless the random phase factors will add-up to 
zero to give a complete destructive interference and leads to Anderson localization. The situation is similar to the Fabry Perrot resonances in disordered 
one dimensional array of alternating dielectric bi layers \cite{makarov}. 
Such a destructive interference related localization takes place 
in the related problems of non-relativistic electrons governed by Schr\"odinger equation.
However for massless two dimensional Dirac fermions 
such a situation is averted because of Klein tunneling, since this ensures that there will be always full transmission at normal incidence. 
\begin{figure}[h]
\begin{center}
\centerline{\epsfxsize 8.5cm \epsffile{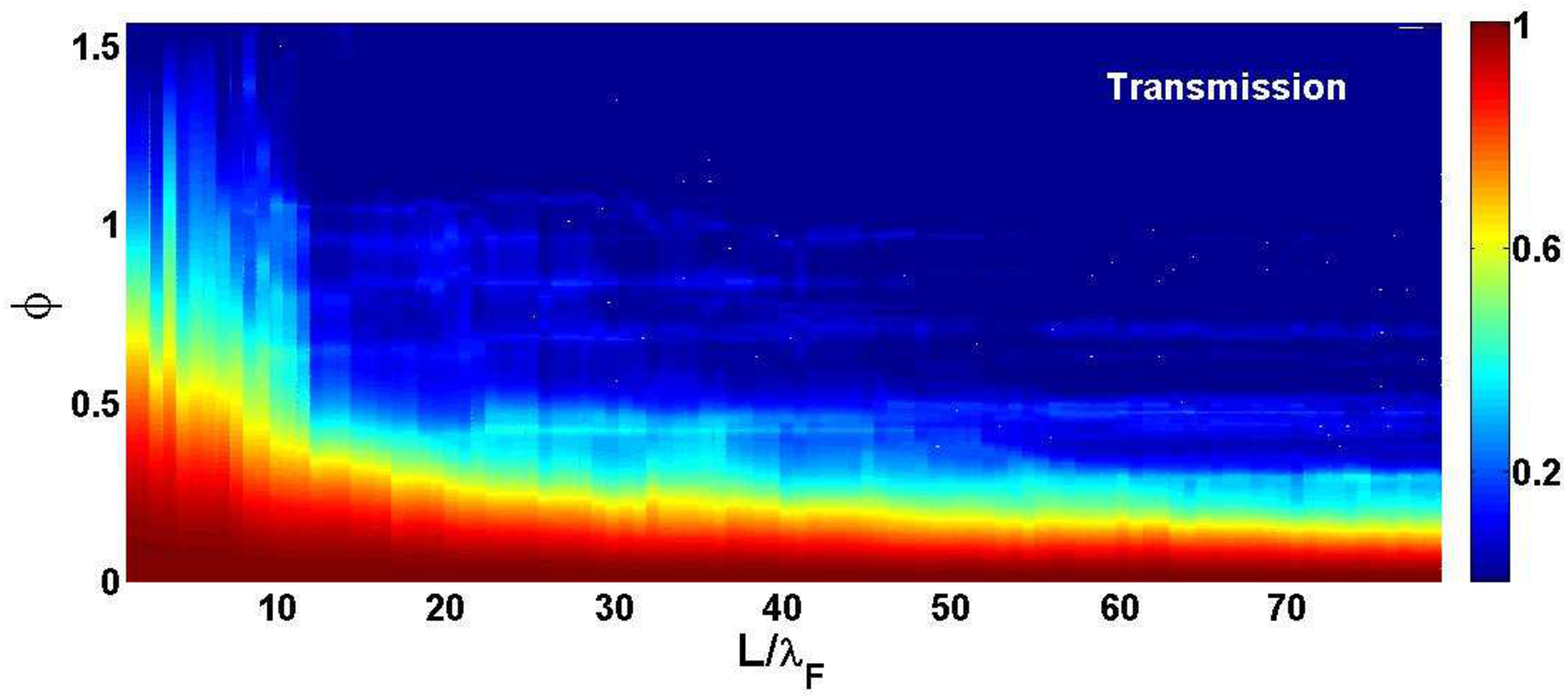}}
\centerline{\epsfxsize 8.5cm \epsffile{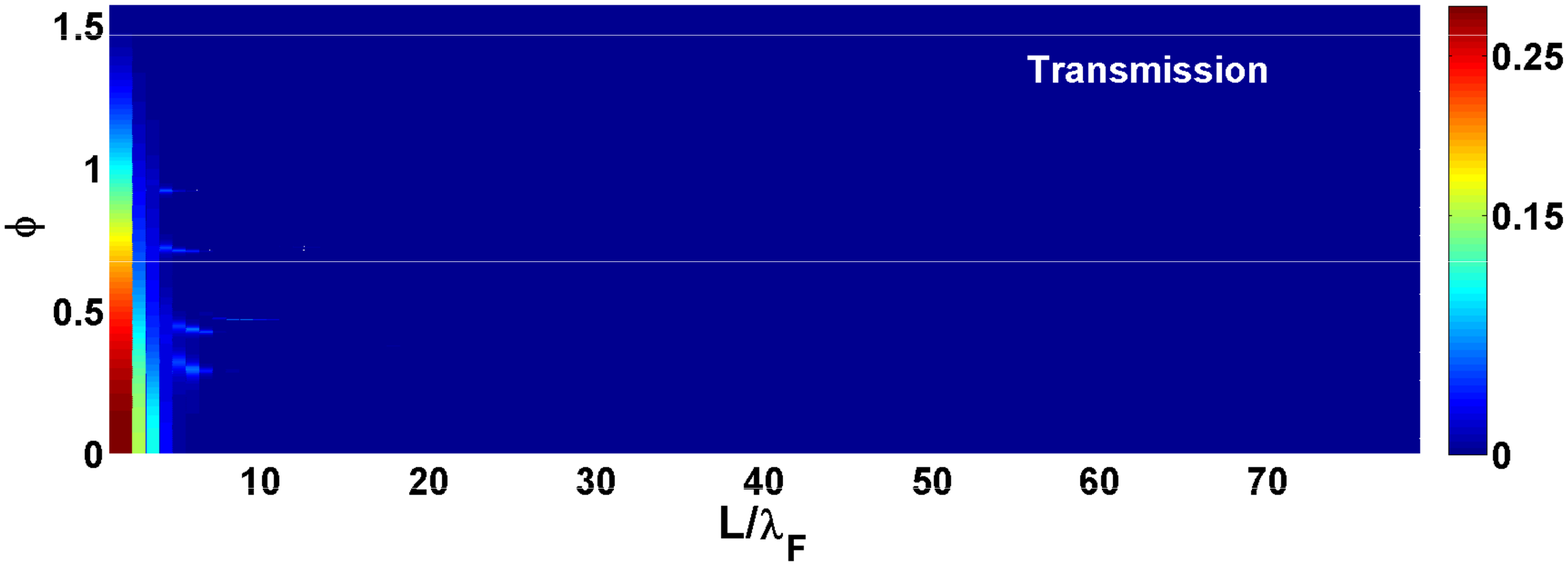}}
\end{center}
\caption{{\it (color online)} Transmission through disorder for (a) Carriers obeying Dirac like equation: The effect of  Klein tunneling. (b) Carriers obeying Schr\"odinger equation. The y axis plotted the angle of incidence. The x-axis plotted
the length of the sample L
 in the unit of Fermi wavelength.}
\label{t}
\end{figure}
The sharp contrast between the transmittance through such one dimensional delta function barriers for two dimensional 
massless dirac fermions and two dimensional non-relativistic electrons obeying Schr\"odinger equation is depicted in 
Fig. \ref{t}. The almost full  transmission around normal incidence for massless dirac fermions can be contrasted again the 
exponential decay of transmission at any angle for non-relativistic electrons.


\begin{figure}[ht]
\begin{center}
\centerline{\epsfxsize 7.5cm \epsffile{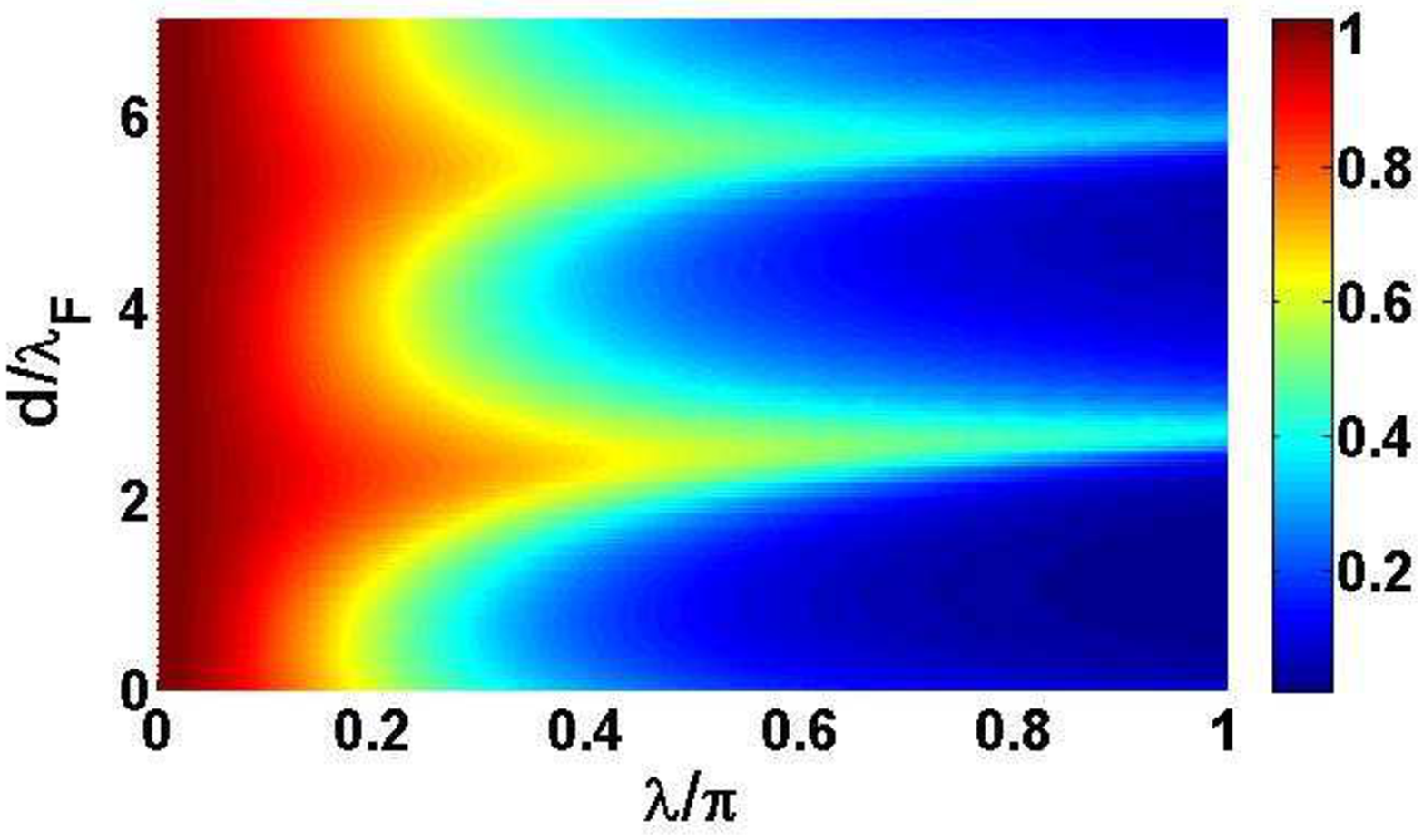}}
\centerline{\epsfxsize 7.5cm \epsffile{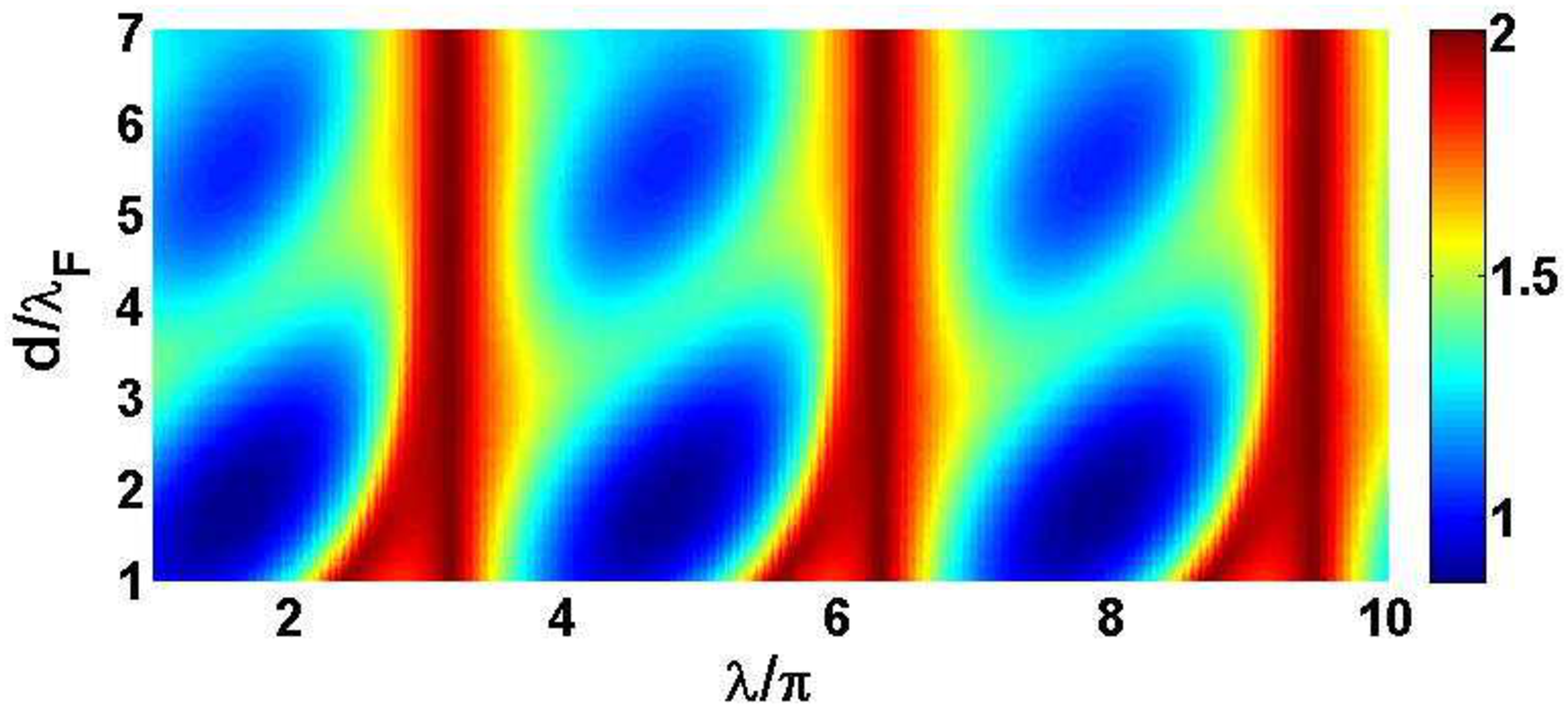}}
\centerline{\epsfxsize 7.5cm \epsffile{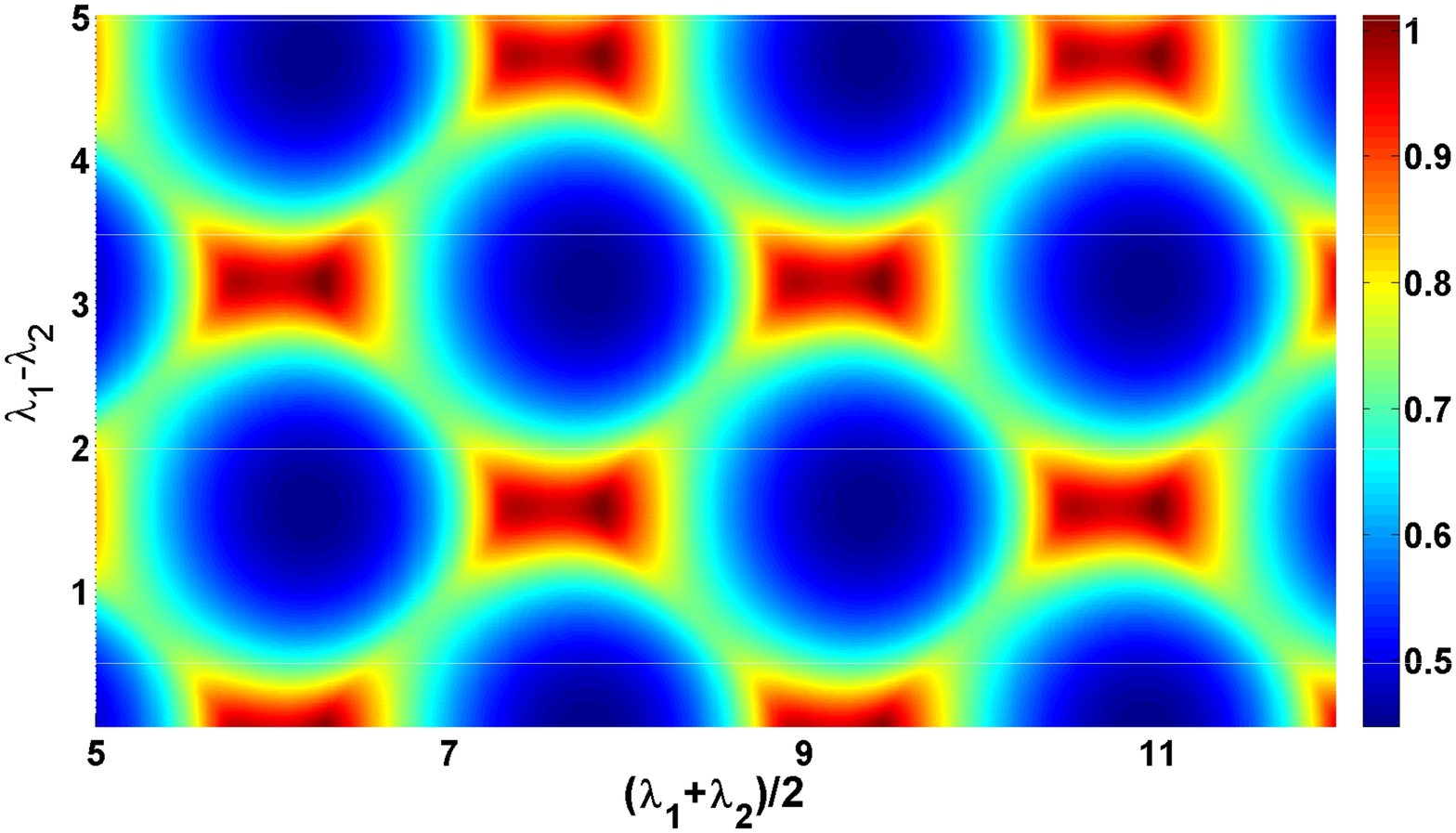}}
\end{center}
\caption{{\it (color online)}Two barrier conductance for non relativistic electrons when the barrier strength are equal. 
(b) Two barrier conductance for graphene charge carriers (b) when the barrier strengths are equal
 (c) Two barrier conductance  when the barrier strengths are unequal. }
\label{dbt}
\end{figure}

The conductance in presence of such double barrier structures is plotted for such massless Dirac fermions in Fig. \ref{dbt} (b) 
and compared against the similar double barrier conductance for non relativistic particles obeying Schr$\ddot{o}$dinger equation 
Fig. \ref{dbt}(a). For the massless dirac fermions, the transmission shows periodicity as function of the separation between 
the barriers ( Fabry Perrot transmission) as well as the strength of the barrier.  The double barrier transmission for the non-relativistic electrons that is depicted in the lower figure shows periodicity as a function of the separation between the barriers only. There is no periodicity as a function of the strength of the barrier for such non relativistic fermions, since 
this is attributed to the ultra relativistic nature of the charge carriers in graphene. 

\begin{figure}[ht]
\begin{center}
\centerline{\epsfxsize 6.5cm \epsffile{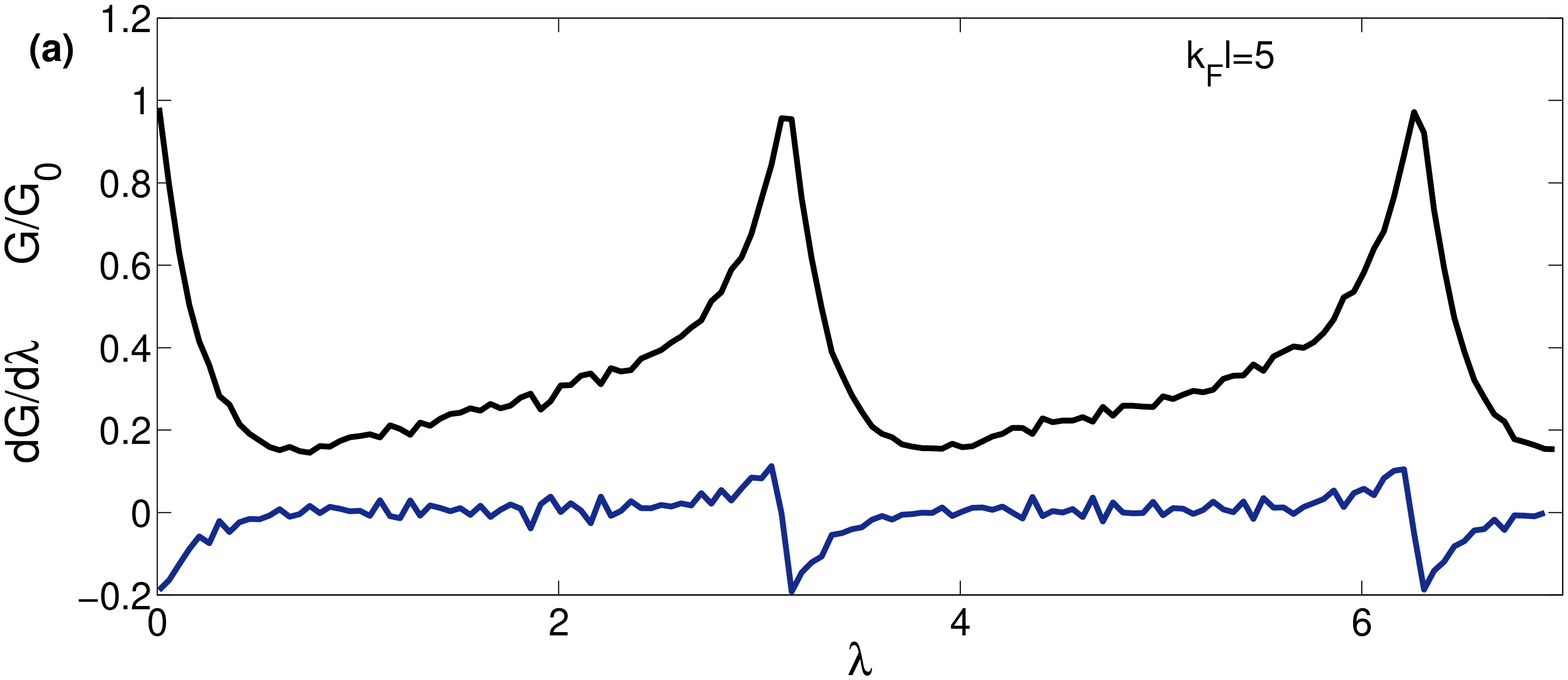}} 
\centerline{\epsfxsize 6.5cm \epsffile{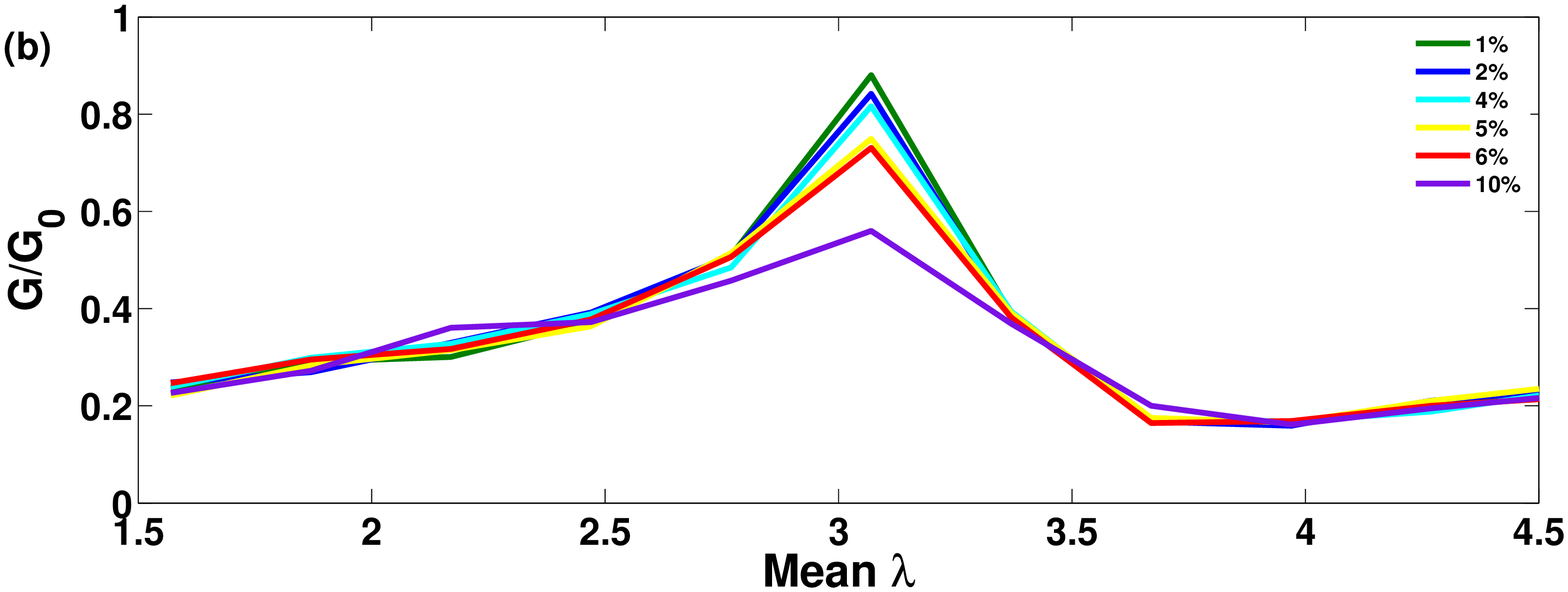}} 
\end{center}
\caption{{\it (color online)} (a) Conductance as a function of strength of potential in the presence of many barriers with position disorder but same potential strength, $\lambda = 2$. Average separation is such that $k_FL = 5$. (b) The change in the shape of conductance resonance as the fluctuation around the mean strength of the barrier is allowed to grow.}
\label{2dperiodic}
\end{figure}

Next we shall consider the transmission of such massless dirac fermions through a double barrier structure when 
the strength of the barriers are unequal, {\it i. e.} $\lambda_{1} \neq \lambda_{2}$. With the help of the expresssion 
(\ref{transformula})  we can evaluate the  resulting conductance which is plotted in  Fig. \ref{dbt} (c) as function of 
the mean strength of the two barriers along the $x$-axis as well as the difference of strength of the two barriers along 
the $y$-axis. Here the amount of pseudopsin rotation imparted by the two barriers are of different. Thus the transmission 
( conductance) resonance occurs when mean $\lambda_{1} + \lambda_{2}$ = 2n $\pi$ and $ \lambda_{1} - \lambda_{2} = 2n \pi$, both the conditions are satisfied.  Also because of the mismatch of the barrier height, the resonace peak has a double hump structure 
which characterizes the resonace due to a double barrier structure with a finite difference between the height of these two barriers. It can be now be 
seen since the transmission through many such barriers with random position and strength 
can  be thought as a product of transmission through such double barrier structure with unequal strength the resonance 
condition for each such pair will be diffferent from another in general and consequently the  height of the conductance 
resonance peak will get reduced with increasing fluctuation around the mean strength of such delta function barriers. 
This fact is demonstrated in Fig. \ref{2dperiodic}. In Fig. \ref{2dperiodic}(a) we have plotted the 
dimensionless conductance in the presence of $N$ randomly positioned barrier all having the same strength.  
As one can see the periodic occurrence of conductance resonance as function of the strength of the delta function barrier.
The lower plot of the same figure shows how differential conductance varies as a function the strength of the delta function. In Fig. \ref{2dperiodic}(b) we plot how this resonance peak changes when apart from the randomness in position we also introduce randomness in strength. In the second case we plot the conductance as a function of the mean strength 
and the fluctuations around this mean. One can see a conductance peak is still observed now at the same mean value 
of $\lambda$ , but with increasing fluctuations around the mean the height of the conductance peak gets reduced.
Current experiments can measure the conductance and differential conductance 
in ballistic regimes  for  graphene based  microstructures \cite{Levitov, kleinexp}. Thus our predictions can be directly verified.  

\subsection{ Transmission and conductance through N barriers}
\subsubsection{Barriers with equal strength}
With the above analysis of transmission through a double delta function like structure, we shall now analyze 
the transmission and the resulting conductance through $N$ such randomly positioned barrier having equal strength $\lambda$. For a given such 
strength, we have varried the length of the sample $L$, mentioned in the unit of Fermi wave vector, $\lambda_{F}$
, keeping the mean separation between the disorder $l$, or the dimensionless quantity $k_{F} l$ constant. Since in all 
our calculations are done for a particular enegy that we call Fermi energy which is in the dimensionless unit $\bar{k}_{F}$,
different $k_{F} l$ correspond to different mean separation between the disorder. 
The higher value of $k_{F} l$ thus implies 
a weakly disordered system where as lower value of $k_{F} l$ implies a relatively strongly disordered system \cite{Ioffe}. 

\begin{figure}
\begin{center}
\centerline{\epsfxsize 8.5cm \epsffile{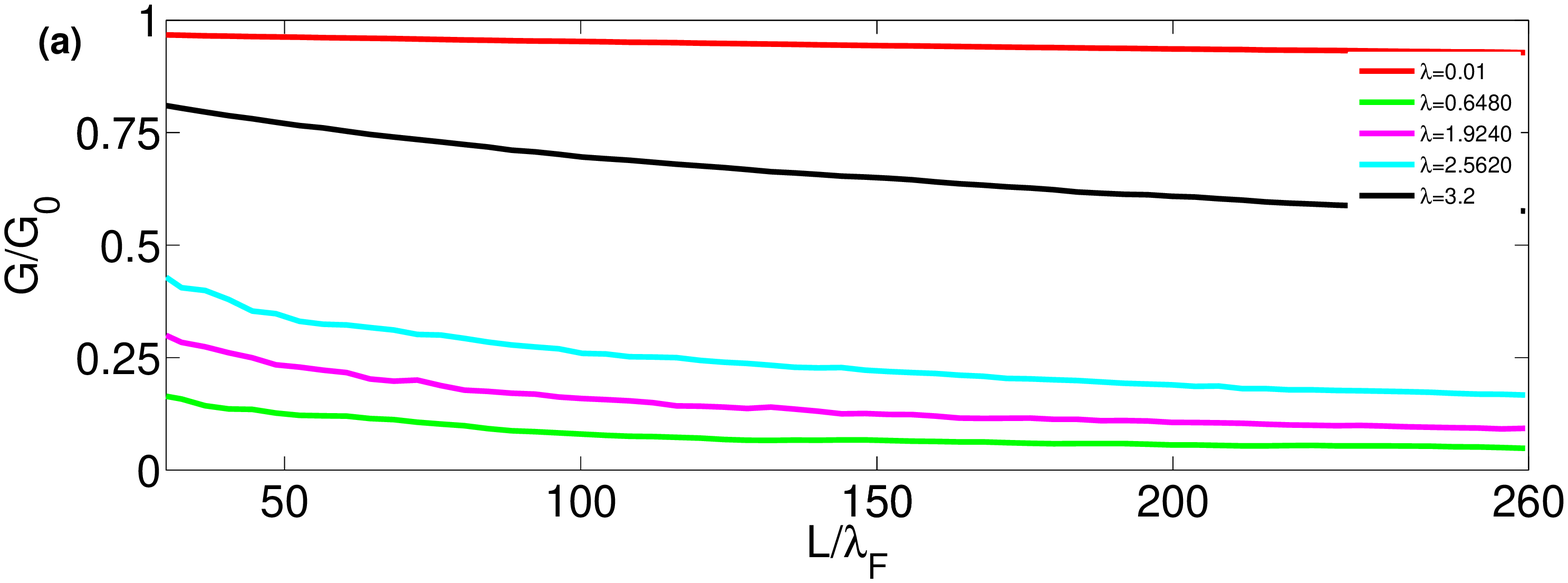}}
\centerline{\epsfxsize 8.5cm \epsffile{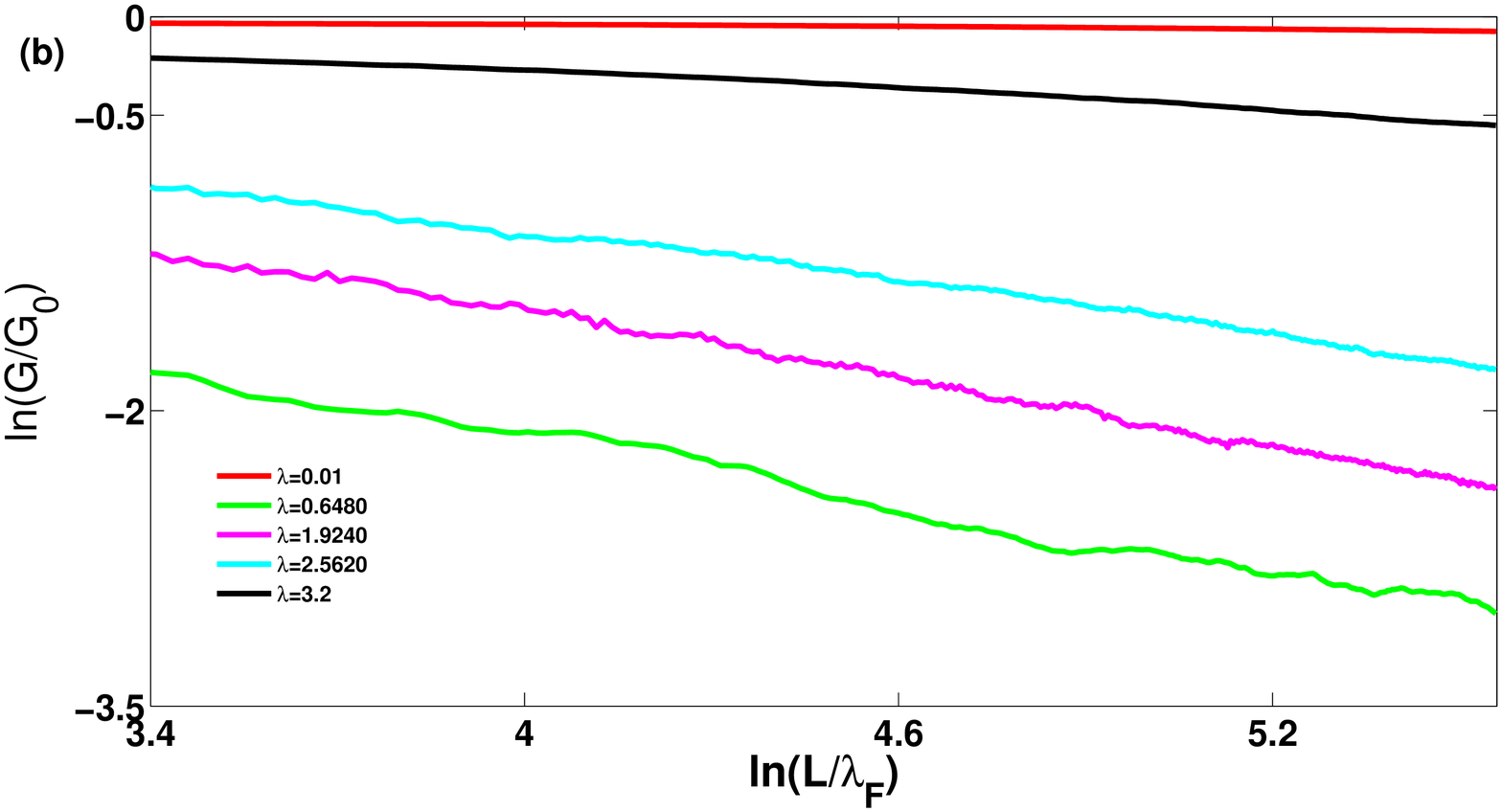}} 
\centerline{\epsfxsize 8.5cm \epsffile{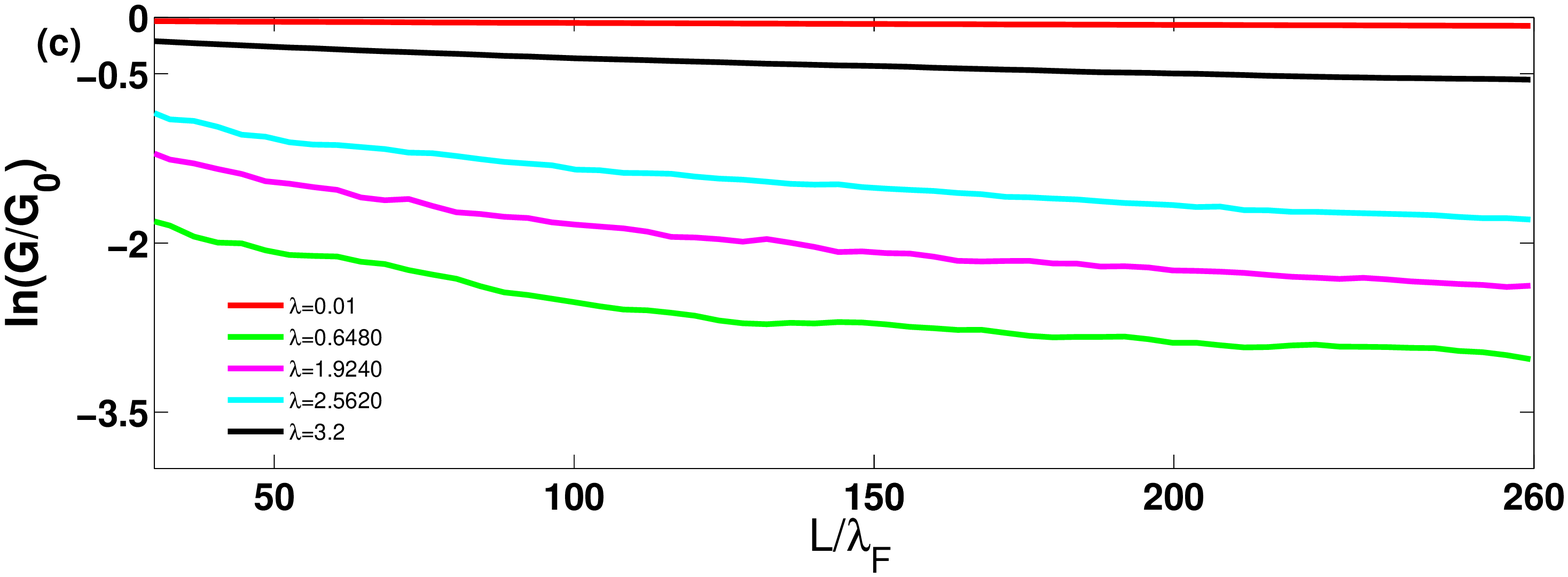}}
\end{center}
\caption{{\it (color online)} Decay of Conductance with increasing sample size, on linear, log-log and semilog scale. Average separation between the barriers is such that $k_FL = 5$, $\epsilon = 1$}
\label{kfl5difflamda}
\end{figure}

\begin{figure}
\begin{center}
\centerline{\epsfxsize 8.5cm \epsffile{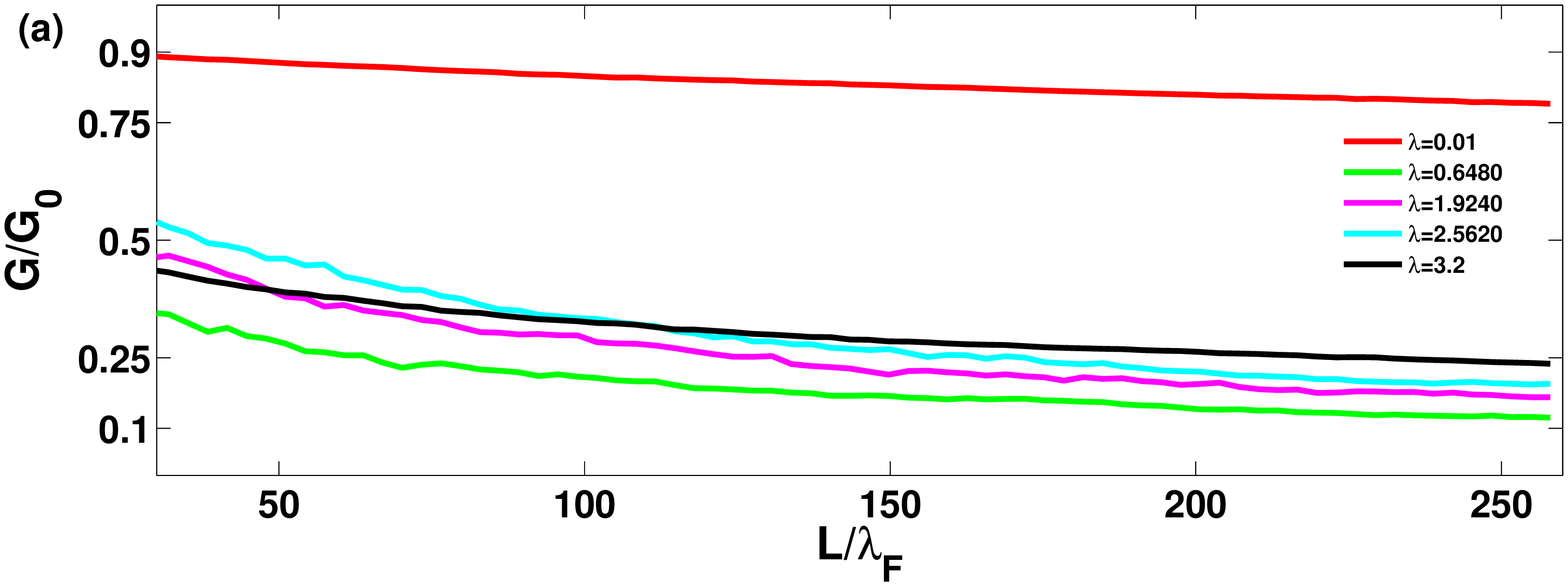}}
\centerline{\epsfxsize 8.5cm \epsffile{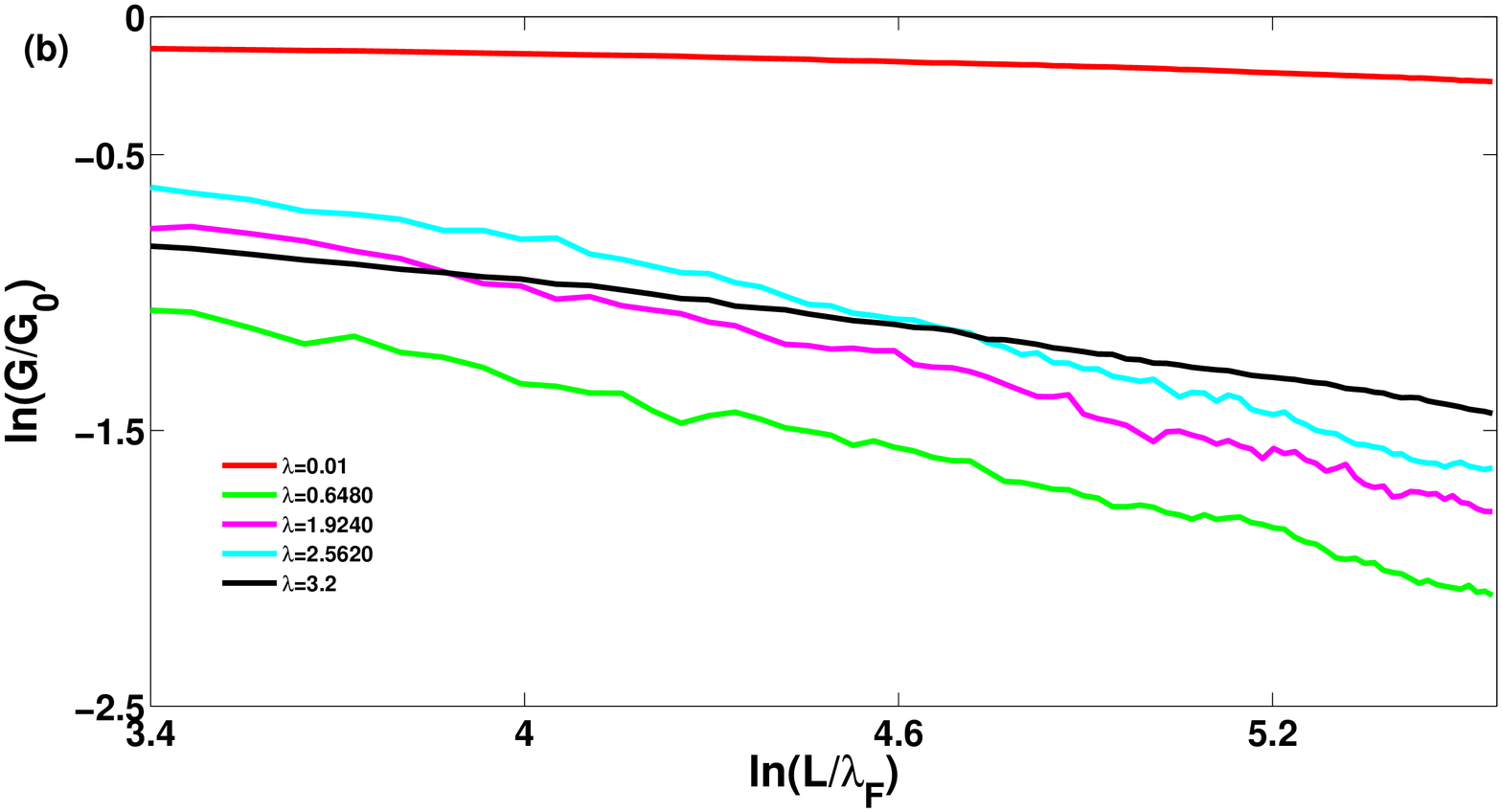}} 
\centerline{\epsfxsize 8.5cm \epsffile{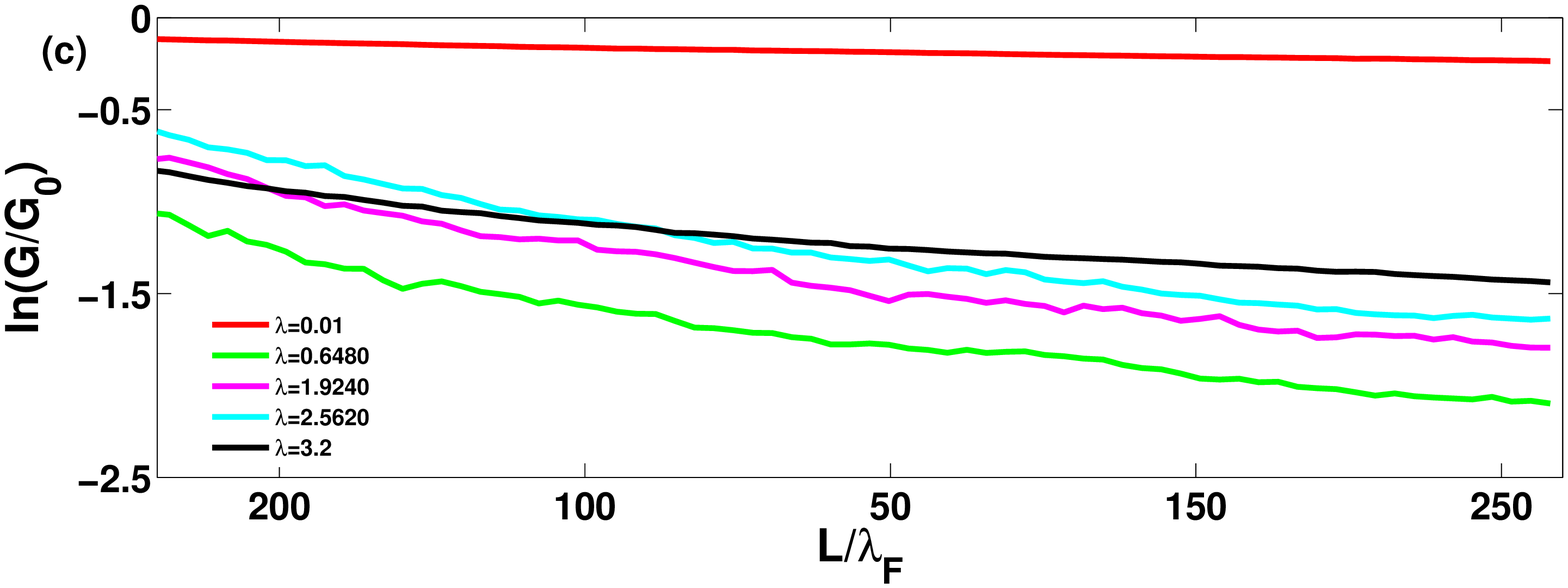}}
\end{center}
\caption{{\it (color online)} Decay of Conductance with increasing sample size, on linear, log-log and semilog scale. Average separation between the barriers is such that $k_FL = 1$, $\epsilon = 1$}
\label{kfl1difflamda}
\end{figure}

With the help of expression (\ref{transformula}) and (\ref{condexp}) we evaluate the dimensionless conductance as a function of the 
sample size, $\frac{G(L)}{G_{0}}$ for different strength of the randomly positioned delta function potentials for a given $k_{F} l$ and a given 
incidence energy $\epsilon$. Such results are presented in Fig. (\ref{kfl5difflamda}) and Fig. (\ref{kfl1difflamda})
for two different values of $k_{F} l$ for a set of disorder or delta function barrier strength $\lambda$. For a given 
llength $L$ ,  $G(L)$ is 
evaluated after doing  ensemble avearge of randomly positioned $N$ delta function like barriers in that length. 

The plot for dimensionless conductance $\frac{G(L)}{G_{0}}$ as a function of $L$ shows 
that for  smaller sample size conductance shows fluctuations. This fluctuations are associated 
with the conductance fluctuations in mesoscopic samples occurs due to the inhomogeneity in the position of the scatterers in such sample and well studied in the literature \cite{Meso}. We shall not discuss this issue further and focus on the behavior 
behavior of $G(L)$ for larger sample size when such fluctuations die down. We found that the $L$ dependence of $G(L)$
can be broadly divided in two parts. For the strength of the deta function barriers satisfying resonant transmission, namely 
$\lambda = n\pi$, $T=1$ and the conductance remains constant as a function of length. Close to this resonant strength, conductance thus show a very slow decay and continuous achieve this resonant behavior. This can be straightforwardly understood with the preceeding discussion of analyzing transmission $N$ random delta function like barrier in terms of two 
barrier transmission.  Away from this resonance strength, the conductance shows an algebraic decay as a function of the sample length $L$. To extract this algebraic decay for different $k_{F} l$ value, we have plotted the $\log \frac{G(L)}{G_{0}}$ as a function of $L$ as well as $\log L$. Fitting these plots we find that for non-resonant strength of the delta 
function barrier $G(L)$ can be well approximated by the expression 
For the non resonant values of strength of potential we obtain:
\beq \frac{G(L)}{G_{0}} = \frac{c}{L^{\alpha}} , \alpha = 0.46-0.56 \label{algebraic} \eeq
with $c$ being a sample dependent constant. 
Our results agrees well with the observation in ref. \cite{titovprl} where a random-matrix theory based argument 
also predicts an algebraic decay of the conductance with exponent $0.5$.

\begin{figure}
\begin{center}
\centerline{\epsfxsize 9cm \epsffile{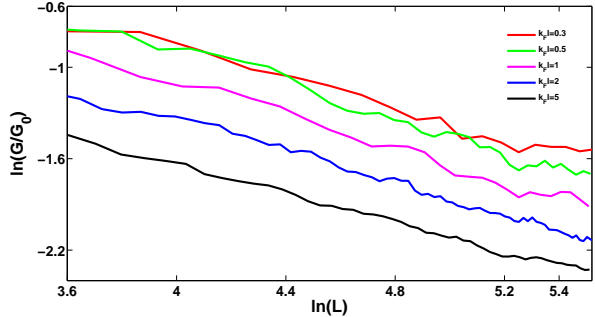}}
\end{center}
\caption{{\it (color online)} Position as well as potential strength disorder: Linear fit on a log-log scale-showing algebraic decay}
\label{fit}
\end{figure}

Finally we plot the conductance on a log-log scale for several values of $k_{F}l$ starting from the $5$ ( weak disorder) 
to $k_{F}l=0.2$ in Fig. \ref{fit}. The smaller values of $k_{F} l$ correspond to fairly strong disorder and one needs to calculate the  ensemble averaged  tranmsission through a very large number of delta function barriers. The fitting of 
of the plots suggest that the algebraic decay again depicts the correct dependence of the conductance on the system size,
when the later is large. However the conductance fluctuationation persists over a larger length scale as the value of $k_{F} l$ is lowered. The much slower decay of the conductance near the resonant value the mean impurity strength also get more and more supressed as suggested by Fig. \ref{2dperiodic} (b) with increasing fluctuations around the mean strength. 

\section{Conclusion}\label{concl}
We conclude by summarizing our main findings. First the  Green's function technique provides a systematic way of understanding the transmission and conductance through short range scatterers in terms of resonant  transport through 
double barrier structure. We particularly point out to distinct regime of transport near the resonant value and off resonant 
values of such short range scatterrs. At and very close to  the resonant value, the conductance in relatively large size sample shows very slow decay as a function of system size where as away from the resonant value the conductance in a large sample shows an algebraic decay as function of the system size with an exponent which is close to $0.5$. Though our 
results are obtained by approximating the short range scatterers as delta function barrier, in order to obtain compact analytical expressions for transmittance and conductance, some of the conclusions can be extended for more extended natured potential barriers as well. 
As our results suggested a transition from this resonant regime to the off resonant regime can be observed by introducing controlled disorder in graphene based superlattice structure
and may well be used to suggest graphene based electronic devices. 

\section{Acknowledgment}
This work is supported by grant SR/S2/CMP-0024/2009 of DST, Govt. of India. One of the authors (NA) acknowledges financial support from CSIR, Govt. of India.


\end{document}